%% file: frascati.tex
\documentclass[10pt,a4paper,twoside,fleqn]{article}  % twocolumn
\usepackage{espcrc2} %{elsevrob}
\usepackage{robpreprintpage}
\usepackage[standard]{refstyle}
\input latexuseful2e.tex

\usepackage[fleqn]{amsmath}
\usepackage{epsfig}

\DFTTtrue
\def\Delphi{\textsc{Delphi}}

\def\myitem#1{\noindent \textit{#1})~}

\title{Oscillations of moments and structure of multiplicity
 distributions in $e^+e^-$ annihilation}
\author{R. Ugoccioni and A. Giovannini\address{
 Dipartimento di Fisica Teorica and I.N.F.N. - Sezione di Torino,\\ 
 via P. Giuria 1, 10125 Torino, Italy}\thanks{Work supported in part by
 M.U.R.S.T. under grant 1996}}

\begin{document}

\setabstract{Starting 
from the recognized fact that oscillations of moments with
rank and shoulder structure in the multiplicity distribution have the
same origin in the full sample of events in \ee\ annihilation, we push
our investigation to the 2-jet sample level,
and argue in favor of the use of the negative
binomial multiplicity distribution as the building block of
multiparticle production in \ee\ annihilation events. 
It will be shown that this approach leads to definite
predictions for the correlation structure, e.g., that correlations
are flavour independent.}

\thispagestyle{empty}

\ifDFTT
\makepreprinttitlepage{DFTT 64/97\\October 10, 1997}{
Oscillations of moments and structure of multiplicity\\
 distributions in $e^+e^-$ annihilation}{R. Ugoccioni and A. 
Giovannini\\[0.3cm]
 \it Dipartimento di Fisica Teorica and I.N.F.N. - Sezione di Torino,\\
 \it via P. Giuria 1, 10125 Torino, Italy}{To be published in the
Proceedings of the\\XXVII International Symposium on Multiparticle
Dynamics\\Frascati (Italy), September 8--12, 1997\\[2cm]
\normalsize Work supported in part by M.U.R.S.T. under grant 1996}\fi

\maketitle
\ifDFTT\else \pagestyle{empty}\fi

\section{FRAMEWORK AND DEFINITIONS} %%----------------------------------

This work is a step in the direction of
the description of multiplicity distributions (MD's) and
correlations functions within a common framework: this
is the approach we chose in order to investigate the 
dynamical mechanism of
multiparticle production.
In particular, in this work we explore the
relationship between the multiplicity distribution, $P_n$, i.e., the
probability of producing $n$ charged final particles, and the
$q$-particle correlation function $C_q(y_1,\dots,y_q)$, using \ee\
annihilation data.

We will use the following moments of the MD:

\myitem{a} the factorial moments:
  \begin{equation}
  F_q = \sum_{n=q}^\infty n(n-1) \cdots (n-q+1) P_n
  \end{equation}

\myitem{b} the factorial cumulant moments:
  \begin{equation}
  K_q = F_q - \sum_{i=1}^{q-1} \binom{q-1}{i} K_{q-i} F_i
  \end{equation}

\myitem{c} their ratio $H_q$:
  \begin{equation}
  H_q = \frac{K_q}{F_q}
  \end{equation}  
and will study these moments as a function of the order $q$ (see
\cite{FaroRU} for more details).
While the
above definitions are valid in any restricted region of phase space,
the work reported here refers only to full phase space.

It should be noticed that the following relation holds:
\begin{equation}
  K_q = \int dy_1 \dots dy_q C_q(y_1,\dots,y_q)
\end{equation}
where the integration is over the full phase space, and
$C_q(y_1,\dots,y_q)$ is the correlation function obtained from the
inclusive $m$-particle densities ($m=1,\dots,q$) by cluster expansion
\cite{DeWolf:rep}.

\begin{figure}[t]
\begin{center}
\mbox{\epsfig{file=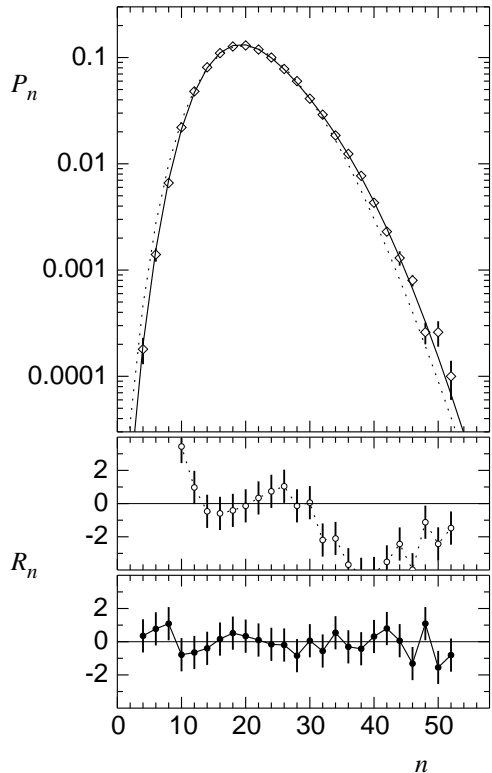,height=10.2cm}}
\end{center}
\vspace{-0.8cm}
\caption[full sample MD]{Final 
charged particles MD for the full sample of events 
at \roots{91} from the \Delphi\ Collaboration (diamonds) and
fits with eq. \ref{eq:fullnbd} (dotted line, open circles
in the residuals) and \ref{eq:full2nbd} (solid line, filled circles
in the residuals).}\label{fig:fullsample}
\end{figure}

\begin{figure}[t]
\begin{center}
\mbox{\epsfig{file=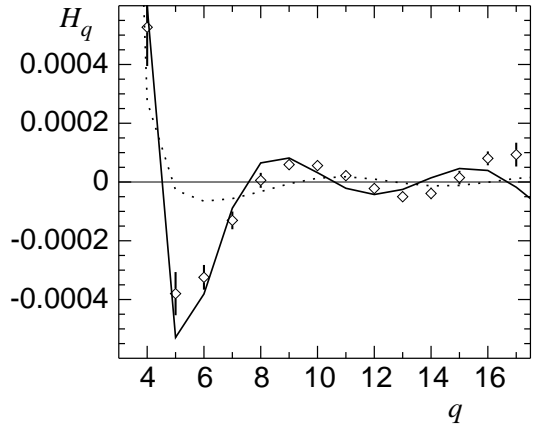,height=5.6cm}}
\end{center}
\vspace{-0.8cm}
\caption{$H_q$ ratio from the  MD's in figure
\protect\ref{fig:fullsample}}\label{fig:fullsamplehq}
\end{figure}

The discussion in the following sections is based on a detailed
analysis of published experimental data. A few points should be made
clear in advance:

\myitem{a} Because of charge conservation, the number of particles in full
phase space  must be even, so all MD's in the following must be
understood as `in their even component only', e.g.,
\begin{equation}
  P_n = P_n^{\text{(NBD)}} (\mu,k)
\end{equation}
should be read as
\begin{equation}
  P_n \propto\begin{cases} 
	P_n^{\text{(NBD)}}(\mu,k)& \text{if $n$ is even}\\
	0                 & \text{if $n$ is odd} 
  \end{cases}
\end{equation}
where a proportionality constant serves the purpose of normalization.
This procedure is normally used in the literature \cite{DeWolf:rep}.

\myitem{b} Because the experimental data sample is finite, the tail
of the distribution cannot be fully sampled, and the published MD's
are indeed truncated at some value; all analytical MD's in the following are
understood to be truncated at the same value as the corresponding
experimental ones. The importance of truncation on the analysis of the
ratio $H_q$ is well known and has been investigated in \cite{hqlett}.

\myitem{c} Fits performed on the published MD's suffer from the fact that
the matrix used to go from the measured MD to the
reconstructed MD in the unfolding procedure
(see, e.g., \cite{DEL:2}) is not usually published.
Thus a correct, unbiased regression cannot be made;
the fits we describe here should then be
seen more as `solid predictions' rather than `solid results'. 
As predictions, they are indeed quite intriguing and our approach
should therefore be considered as a strong suggestion to the
experimental groups.

\myitem{d} For the same reason discussed in \textit{c)}, the error
bars we plot here for the ratio $H_q$ should be considered indicative:
the method of varying randomly each $P_n$ within 
its standard error with a Gaussian distribution many times in order to
calculate the variance of the $H_q$ should be
applied to the uncorrelated $P_n$ (i.e., before unfolding); however, 
even if we apply it to published data, it
agrees with experimental results \cite{hqlett:2}.

\section{THE FULL SAMPLE OF EVENTS} %%--------------------------------

We begin by examining the full sample of events at \roots{91}.
Data on MD's have been published by all Collaborations, but in order
to be consistent and concise we will only use the data
published by the \Delphi\ Collaboration \cite{DEL:2} 
(reference \cite{hqlett:2} contains a more detailed analysis). 
In \cite{DEL:2}, the negative binomial distribution (NBD) is fitted
to the data;
it is worth recalling the explicit expression for the NBD:
\begin{multline}
\hspace{-0.3cm}P_n^{\text{(NBD)}}(\mu,k) =\\ \frac{k(k+1)\cdots(k+n-1)}{n!}
	\frac{ \mu^n k^k}{ (\mu+k)^{n+k} }  \label{eq:fullnbd}
\end{multline}
where the two parameters $\mu$ and $k$ are related to the first two
moments as 
\begin{equation}
  \mu = \avg{n} \quad\text{and}\quad  k^{-1} = K_2/\mu^2 \label{eq:muk}
\end{equation}
The description of the data by the NBD
is clearly unsatisfactory from the point of view of
the chi-square (see table \ref{tab:params}a), 
and also from the point of view of the residuals
(shown in figure \ref{fig:fullsample} with open dots): there is a
clear structure which relates to what has been called 
`shoulder structure' in $P_n$ \cite{DEL:4}.
The corresponding situation with the ratio $H_q$ is shown in
figure \ref{fig:fullsamplehq}, from which it is evident 
that a single truncated NBD (dotted line) cannot
describe the oscillations of $H_q$ with $q$.

It was shown by the \Delphi\ Collaboration  \cite{DEL:4}
that the shoulder structure
can be explained in terms of the superposition of the MD's resulting from 
classifying the events according to the number of jets, 
and that the MD in each class is well
fitted by a NBD. Moreover, the goodness of this description is not
influenced by varying the parameter of the jet-finding algorithm.

It is then natural to try a description of the full sample of events
with the weighted sum of two NBD's \cite{hqlett:2}, 
one corresponding to 2-jet events
and the other to 3-jet events (4-jet events' contribution is negligible).
This is implemented by forcing the weight to be equal to the fraction
of 2-jet events as experimentally measured in \cite{DEL:4}. 
The resulting MD is thus
\begin{multline}
\hspace{-0.3cm}P_n = \alpha_{\text{2-jet}} 
	  P_n^{\text{(NBD)}}(\mu_{\text{2-jet}},k_{\text{2-jet}}) + \\
	(1-\alpha_{\text{2-jet}}) 
	  P_n^{\text{(NBD)}}(\mu_{\text{3-jet}},k_{\text{3-jet}}) 
					\label{eq:full2nbd}
\end{multline}
This fit is shown with a solid line in figure \ref{fig:fullsample}
for $\alpha_{\text{2-jet}} = 0.767$, corresponding to $y_{\text{min}}
= 0.06$ in the jet-finder JADE;
the parameters of the fit are shown in table \ref{tab:params}b.
In the same figure we also notice that the structure of
the residuals has greatly improved, and in figure \ref{fig:fullsamplehq}
that the $H_q$ ratio is also well described (solid line). 
Furthermore, the parameters we find from
our fit are consistent with those obtained experimentally fitting the
2- and 3-jet samples separately with single NBD's.
Analogous result can also be obtained analyzing the SLD and OPAL data
\cite{hqlett:2}; L3 data \cite{L3:hq} are consistent with SLD's.

\begin{table}
\caption{Parameters of the MD's fitted to the data
as shown in the figures. Part a) refers to figure \ref{fig:fullsample}
(dotted line); part b) to figure \ref{fig:fullsample} (solid line);
part c) to figure \ref{fig:2jetsample} (dotted line);
part d) to figure \ref{fig:2jetsample} (solid
line).}\label{tab:params}
% make '?' active and 1 digit wide:
% adapted from TeX book, p. 241
\newlength{\digitwidth} \settowidth{\digitwidth}{\rm 0}
\catcode`?=\active \def?{\kern\digitwidth}
\begin{center}
\begin{tabular}{rlc}
\hline
a) & $\mu$  & $21.4 \pm 0.9$ \\
   & $k$    & $24.3 \pm 0.7$ \\
   & $\chi^2$/NDF &  ?80/34\\
\hline
b) & $\mu_{\text{2-jet}}$   & $19.4 \pm 0.2$ \\
   & $k_{\text{2-jet}}$     & $52 \pm 6$ \\
   & $\mu_{\text{3-jet}}$   & $27.3 \pm 0.3$ \\
   & $k_{\text{3-jet}}$     & $?61 \pm 10$ \\
   & $\chi^2$/NDF & ?12/21 \\
\hline
c) & $\mu$  & $18.5 \pm 0.1$ \\
   & $ k$   & $57 \pm 3$ \\
   & $\chi^2$/NDF & ?20/16 \\
\hline
d) & $\mu_{\text{light}}$  & $17.2 \pm 0.2?$ \\
   & $\mu_{\text{heavy}}$  & $22.0 \pm 1.6? $ \\
   & $ k$     & $145 \pm 53? $ \\
   & $\chi^2$/NDF & 13/16 \\
\hline
\end{tabular}
\end{center}
\end{table}

\section{THE 2-JET SAMPLE OF EVENTS} %%-----------------------------

\begin{figure}[t]
\begin{center}
\mbox{\epsfig{file=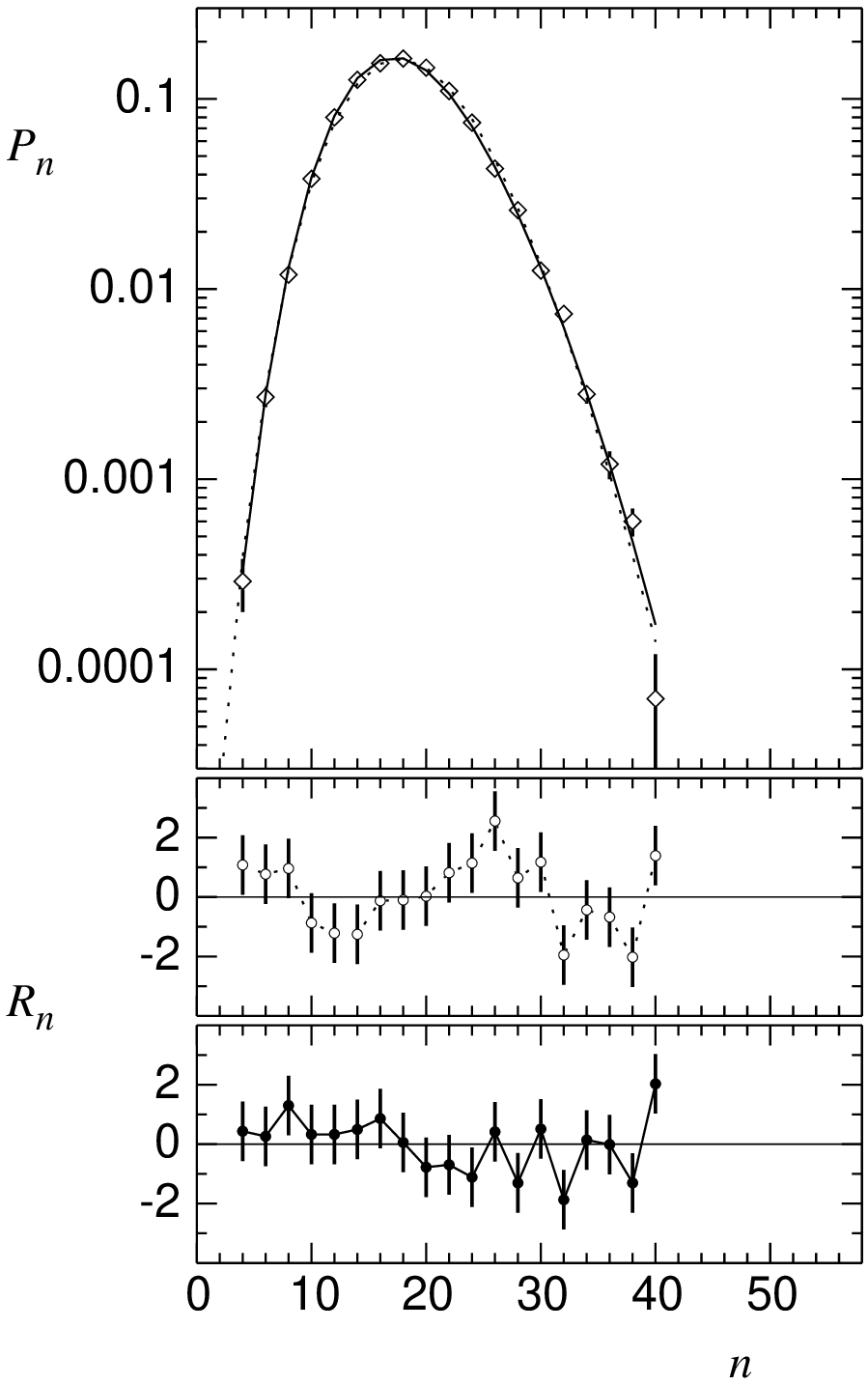,height=10.2cm}}
\end{center}
\vspace{-0.8cm}
\caption[full sample MD]{Final 
charged particles MD for the 2-jet sample 
of events ($y_{\text{min}} = 0.02$)
at \roots{91} from the \Delphi\ Collaboration (diamonds) and
fits with eq. \ref{eq:fullnbd} (dotted line, open circles
in the residuals) and \ref{eq:nbdk} (solid line, filled circles
in the residuals).}\label{fig:2jetsample}
\end{figure}

\begin{figure}[t]
\begin{center}
\mbox{\epsfig{file=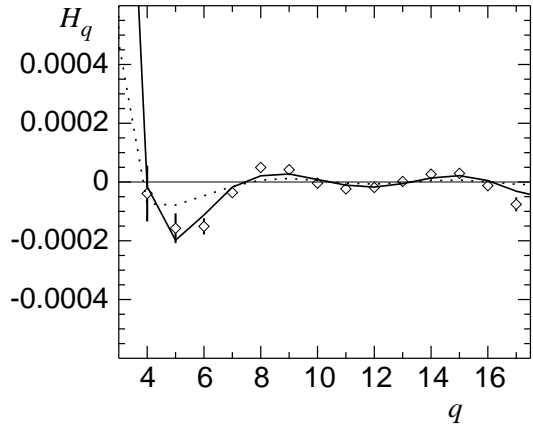,height=5.6cm}}
\end{center}
\vspace{-0.8cm}
\caption{$H_q$ ratio from the MD's in figure
\protect\ref{fig:2jetsample}}\label{fig:2jetsamplehq}
\end{figure}

It is a striking result, and a lesson to be learned from the analysis 
mentioned in the previous section,
that one can resolve puzzling structures by investigating them on a
deeper level. It is a prediction of the same analysis that the 2-jet
sample should be well described (in all its aspects) by one NBD.
However, this turns out not to be entirely true,
as shown in figure \ref{fig:2jetsample},
which reproduces the 2-jet data and (single) NBD fit by \Delphi\
\cite{DEL:4}. While the chi-square is satisfactory (see table
\ref{tab:params}c), the residuals (open
dots in the figure) show some structure, although much less pronounced
than in the full sample; the ratio $H_q$ 
(figure \ref{fig:2jetsamplehq}) still shows oscillations,
albeit an order of magnitude less deep than in the previous case.

A few results coming from experiments will help us investigate
deeply the structure of 2-jet events.
First, the result by OPAL \cite{OPAL:FB} on forward--backward
correlations: they are weak, and 
in the bulk they are explained by the superposition of events with a
fixed number of jets; the residual correlation is attributed to
the presence of events with quarks of heavy flavour. This suggests
that longitudinal observables can reveal the flavour structure of the
original hard event.
Second, the result by \Delphi\ on the MD in one hemisphere \cite{DEL:bb}: 
a sample enriched in $b\bar b$ events was found to
be essentially identical in shape to the MD of the full sample, apart
from a shift of one unit. If taken literally, and if forward--backward
correlations are neglected except for the requirement of even total
charged multiplicity, this result suggests that the $b\bar b$ events
MD, when the fraction of $b\bar b$ events is small,
is very similar to the {\em light} $q\bar q$ MD, except for a translation
of two units.

Guided by all the above considerations, we propose to use a NBD to
describe the MD of a 2-jet event of fixed flavour. We identify two
types of hadronic events: those originating from a $b\bar b$ pair 
and those from a  $q\bar q$ pair, 
where with $q$ we indicate all flavours lighter than $b$. We thus
use for the MD of 2-jet events the weighted sum of two NBD's:
\begin{multline}
\hspace{-0.3cm}P_n = \alpha_{\text{heavy}} 
	   P_n^{\text{(NBD)}}(\mu_{\text{heavy}},k) + \\
	(1-\alpha_{\text{heavy}}) P_n^{\text{(NBD)}}(\mu_{\text{light}},k)  
				\label{eq:nbdk}
\end{multline}
Here $\alpha_{\text{heavy}} = 0.22$ is 
the experimentally determined fraction of $b\bar b$
decays of the $Z^0$ at \roots{91}. The subscripts `heavy' and `light' refer
respectively to $b\bar b$ and $q\bar q$ events. It should be noticed
that the parameter $k$ is the same in both NBD's, while the difference
between the average values is not fixed. This reflects the
above considerations on single hemisphere multiplicity.

The description of published data by the above parameterization is
very good, as shown in  table
\ref{tab:params}d and in figures \ref{fig:2jetsample} and
\ref{fig:2jetsamplehq} (solid lines and solid dots), 
from all points of view: chi-square, residuals and
$H_q$ analysis. It is further supported by the fact that a few
simpler alternatives do not work as well \cite{hqlett:3}.

\section{SUMMARY AND CONCLUSIONS} %%---------------------------------

A relatively simple parameterization, based on phenomenological
considerations 
and previous experimental analyses and results, has been shown to
describe very well the charged particle multiplicity distribution in
\ee\ annihilation at LEP energy, including the tail, that is the
moments of high order. This is so even at the very specific level
of the 2-jet sample of events.

This parameterization is characterized by the
presence of two types of events (heavy and light quarks decays of the
$Z^0$) and by the fact that each of these is described by a negative
binomial distribution. Since for 2-jet events each hemisphere
correspond to one jet, and the two hemisphere are practically
uncorrelated, we conclude that the MD of a ``single quark-jet'' is
described by the NBD. Thus we are lead to suggest that the NBD is the
fundamental building block of multiparticle production in \ee\
annihilation.

Finally, it was shown that the single quark-jet NBD has a parameter
$k$ which is the same for all flavours. Because of the relation
between $k$ and $K_2$, eq. \ref{eq:muk} above, we conclude that true
correlations are expected to be flavour independent.

It should be stressed that all mentioned conclusions are
predictions which can easily be the
subject of experimental testing;
we think indeed that the results of these tests
will improve our understanding of multiparticle dynamics
and of the properties of strong interactions.

\raggedbottom
\pagebreak[4]
\section*{ACKNOWLEDGEMENTS}

The work reported here was in done in collaboration with Sergio Lupia
(MPI, %(Max-Planck-Institut, 
M\"unchen).

We would like to thank Giulia Pancheri and all the organizers of this very
successful symposium for the nice atmosphere that was created.

\input{frascati.ref}

\end{document}

%% file: latexuseful2e.tex
\def\ee{$e^+e^-$}               % e+ e- (annihilations...)

\newcommand{\roots}[1]{$\sqrt{s} = #1$ GeV} % the well abused root s = xx GeV

\def\pt{p\kern -.2pt\lower 4pt\hbox{\tiny T}}    %works?

\def\p0{P_0(\Delta y)}

\def\avg#1{\langle #1 \rangle}  % <n>:

\def\NF{\mathcal{N}_{\kern -1.9pt f}}
\def\NC{\mathcal{N}_{\kern -1.7pt c}}

%% file: frascati.ref
%%%% References file: frascati.ref
%%%% Creator: REFLATEX.PL 1.0
%%%% Usage: \input{frascati.ref} where references should go.
%%%%